%% file: paper.tex
\documentclass{article}

\usepackage{microtype}
\usepackage{graphicx}
\usepackage{subfigure}
\usepackage{booktabs} 

\usepackage{hyperref}

\usepackage{xspace}
\usepackage{multirow}
\usepackage{diagbox}
\usepackage{xcolor}
\usepackage{caption}
\def\etc{etc.\@\xspace}
\def\ie{i.e.,\@\xspace}
\DeclareRobustCommand\onedot{\futurelet\@let@token\@onedot}
\def\onedot{\ifx\@let@token.\else.\null\fi\xspace}
\def\eg{\emph{e.g}\onedot}

\usepackage[accepted]{icml2023}

\usepackage{amsmath}
\usepackage{amssymb}
\usepackage{mathtools}
\usepackage{amsthm}

\usepackage[capitalize,noabbrev]{cleveref}

 \def\compactify{\itemsep=0pt \topsep=0pt \partopsep=0pt \parsep=0pt}
\newcommand{\compress}{\itemsep=0pt \topsep=0pt \partopsep=0pt \parsep=0pt \leftmargin=30pt \labelwidth=30pt}

\let\latexusecounter=\usecounter

\newenvironment{CompactEnumerate}
  {\def\usecounter{\compactify\latexusecounter}
   \begin{enumerate}}
  {\end{enumerate}\let\usecounter=\latexusecounter}

\theoremstyle{plain}

\theoremstyle{definition}

\theoremstyle{remark}

\usepackage[textsize=tiny]{todonotes}

\icmltitlerunning{Scalable Multi-Agent Reinforcement Learning through Intelligent Information Aggregation}

\begin{document}

\twocolumn[
\icmltitle{Scalable Multi-Agent Reinforcement Learning through\\ Intelligent Information Aggregation}



\icmlsetsymbol{equal}{*}

\begin{icmlauthorlist}
\icmlauthor{Siddharth Nayak}{mit}
\icmlauthor{Kenneth Choi}{mit}
\icmlauthor{Wenqi Ding}{mit}
\icmlauthor{Sydney Dolan}{mit}
\icmlauthor{Karthik Gopalakrishnan}{stan}
\icmlauthor{Hamsa Balakrishnan}{mit}
\end{icmlauthorlist}

\icmlaffiliation{mit}{Massachusetts Institute of Technology, Cambridge, USA}
\icmlaffiliation{stan}{Stanford University, Stanford, USA}

\icmlcorrespondingauthor{Siddharth Nayak}{sidnayak@mit.edu}

\icmlkeywords{multi-agent reinforcement learning, graph neural networks, coordination and control}

\vskip 0.3in
]



\printAffiliationsAndNotice{}  

\begin{abstract}
We consider the problem of multi-agent navigation and collision avoidance when observations are limited to the local neighborhood of each agent. We propose InforMARL, a novel architecture for multi-agent reinforcement learning (MARL) which uses local information intelligently to compute paths for all the agents in a decentralized manner. Specifically, InforMARL aggregates information about the local neighborhood of agents for both the actor and the critic using a graph neural network and can be used in conjunction with any standard MARL algorithm. We show that (1) in training, InforMARL has better sample efficiency and performance than baseline approaches, despite using less information, and (2) in testing, it scales well to environments with arbitrary numbers of agents and obstacles. We illustrate these results using four task environments, including one with predetermined goals for each agent, and one in which the agents collectively try to cover all goals. Code available at \href{https://github.com/nsidn98/InforMARL}{https://github.com/nsidn98/InforMARL}.
\end{abstract}

\input{sections/1_introduction}
\input{sections/2_related_work}
\input{sections/3_methodology}

\input{sections/4_experiments}

\input{sections/5_conclusion_future}
\input{sections/6_acknowledgement}

\bibliography{references}
\bibliographystyle{icml2023}

\newpage
\appendix
\onecolumn

\input{sections/7_appendix} 

\end{document}

%% file: sections/1_introduction.tex
\section{Introduction}
Reinforcement Learning (RL) has seen wide-ranging successes recently in high-dimensional robot control \cite{DDPG, RLVisuomotor}, solving physics-based control problems \cite{contControlRL}, playing Go \cite{alphaGo}, Chess \cite{alphaZero} and Atari video games \cite{DQN_Atari, Atari2}, \etc However, challenges remain in many real-world applications in which the tasks cannot be handled by a single agent, e.g., multi-player games, search-and-rescue drone missions, \etc \cite{MARLSurvey}. In such cases, multiple agents may need to work together and share  information in order to accomplish the task \cite{MARLseminal}. Na{\"i}ve extensions of single-agent RL algorithms to multi-agent settings do not work well because of the non-stationarity in the environment, \ie the actions of one agent affect the actions of others \cite{IQL1, IQL2}. Furthermore, tasks may require cooperation among the agents. Classical approaches to optimal planning may (1) be computationally intractable, especially for real-time applications, and (2) be unable to account for complex interactions and shared objectives between multiple agents. The ability of RL to learn by trial-and-error makes it well-suited for problems in which optimization-based methods are not effective. In particular, multi-agent reinforcement learning (MARL) approaches may be suitable in these situations due to their fast run-times and superior performance, and their ability to model shared goals between agents using appropriate reward structures.

In this paper, we focus on multi-agent navigation and collision avoidance problems in which there are $N$ agents trying to cooperate with each other to collectively solve a task in a 2D environment with static and/or dynamic obstacles. 
The rewards are shared across all agents in these collaborative environments. We assume that an agent can only sense the presence of obstacles or other agents within a certain limited radius $r$. 
The overarching objective is for all the agents to complete their tasks in the shortest time possible, while avoiding collisions with other agents and obstacles.
This problem setting is quite general and arises in many contexts, \eg, search-and-rescue robot teams \cite{search_rescue}, environmental monitoring \cite{env_monitoring}, and drone delivery systems \cite{drone_delivery1, drone_delivery2, drone_delivery3}.

MARL-based techniques have achieved significant successes in recent times, \eg, DeepMind's AlphaStar surpassing professional level players in StarCraft II \cite{AlphaStar}, OpenAI Five defeating the world-champion in Dota II \cite{Dota2}, etc. The performance of many of these MARL algorithms depends on the amount of information included in the state given as input to the neural networks \cite{MAPPO}.
In many practical multi-agent scenarios, each agent aims to share as little information as possible to accomplish the task at hand. This structure naturally arises in many multi-agent navigation settings, where agents may have a desired end goal but do not want to share their information due to communication constraints or proprietary concerns \cite{satelliteInfoSharing1, satelliteInfoSharing2}. These scenarios result in a decentralized structure, as agents only have locally available information about the overall system's state. In this paper, we focus on the question: ``\textit{Can we train scalable multi-agent reinforcement learning policies that use limited local information about the environment to perform collision-free navigation effectively?}" 

We propose \textit{InforMARL}, an approach for solving the multi-agent navigation problem using graph-reinforcement learning. The main features of our approach are that it: (1) uses a graph representation of the navigation environment which enables local information-sharing across the edges of the graph; (2) transfers well to different numbers of agents; and (3) achieves better sample complexity in training compared to prior approaches by aggregating relevant local information from neighbors in the underlying graph. More broadly, our work (1) demonstrates that graphs provide a valuable abstraction for multi-agent navigation environments; (2) highlights that more information (\ie global information as states) may not necessarily improve performance, and can, in fact, overwhelm the RL agent networks and lead to increased sample complexity; and (3) shows how graph architectures can identify the most valuable information for navigation from local observations to improve performance and scalability.


%% file: sections/2_related_work.tex
\section{Related Work}
%
\label{section:related_work}
Multi-agent navigation problems arise in a number of other contexts, \eg multi-robot navigation. Optimization-based \cite{RVO} and trajectory-based \cite{SIPP,probSafeNav,trajPlan} approaches have been used for multi-robot navigation. However, the former often has long computation times making real-time execution infeasible, while the latter can encounter the ``freezing robot problem" in dense environments due to a large portion of the statespace being marked as unsafe. We therefore focus on MARL approaches. 
\subsection{Scaling MARL}
Research on scaling MARL algorithms has broadly followed two main themes: (1) decentralized execution, and (2) transferring learning between scenarios.

\textbf{Decentralized MARL}:
Centralized-training-decentralized-execution (CTDE) is a popular approach to improve scaling. CTDE frameworks typically use actor-critic methods \cite{ActorCriticSeminal}, where the training step uses a centralized critic that incorporates global information from all actors. During execution, the agents use their own actor networks to select their actions in a decentralized manner. 
MADDPG \cite{MADDPG} builds upon DDPG \cite{DDPG} by learning a centralized critic that is provided the joint state and actions of all agents. MATD3 \cite{MATD3} uses a double-centralized critic model, reducing the over-estimation bias. \cite{MAPPO} show the effectiveness of PPO in several standard multi-agent environments. VDN \cite{VDN} decomposes a centralized value function to a sum of individual agent-specific functions. Q-Mix \cite{QMIX} improves upon this by imposing a monotonicity requirement on agents' individual value functions and using a learnable mixing of the individual functions.  These MARL algorithms perform well in the navigation environment when they know the positions of all entities in the environment. But, as we will demonstrate, they fail to learn when that information is restricted to just local neighborhoods around the agents.



\textbf{Transferability in MARL}:
It is desirable to have MARL formulations where the number of agents/entities in the environment doesn't hinder the performance of the model. Most of the previous works using MARL for the navigation task require concatenation of observations of other entities in the environment to be able to learn meaningful policies. As the neural network size depends on the state input dimensions used while training, the learned policy fails to work in scenarios with a different number of entities in the environment. With the recent success of graph neural networks, many recent works have focused on leveraging the inherent graph structure present in multi-agent problems and tackling the issue of transferability. Zhou et al. \cite{GNN_IL} create a neighborhood graph according to how close the entities are to each other and use imitation learning to imitate a greedy behavior for the target-coverage problem \cite{targetCoverage1, targetCoverage2}. Similarly, Khan et al. \cite{GPG} use graph neural networks with vanilla policy gradient \cite{policyGradient} for the formation flying task \cite{formationFlying1, formationFlying2}. They show that dynamic graphs have worse performance than static graphs due to the large number of possible graphs the model has to learn. At the beginning of each episode, their model determines the connectivity of the agents with each other and uses the same graph over the whole episode. This works well for the formation flying task as the graph structure does not change much if the agents fly in the same formation over an episode. 
DGN \cite{DGN} combines a graph convolutional neural network \cite{GCN} architecture with multi-head attention \cite{AttentionIsAllYouNeed} for a variety of multi-agent environments, including 2D coverage and tracking.
The DGN architecture assumes that each agent communicates with its three closest neighbors. Communicating with the three closest agents leads to a graph that is always connected. However, the three-agent connectivity assumption is highly restrictive in real life applications, as communication is generally restricted by mutual separation due to hardware constraints. Similarly, \cite{G2ANet, GNN_CAV} use an equivalent DGN-like architecture for multi-agent control.

We use a distance-based agent-entity graph similar to Entity-Message Passing (EMP) \cite{EMP}. However, the EMP architecture assumes that the agents have access to the positions of all entities in the environment at the beginning of the episode, which is not possible if there are occluded static or dynamic obstacles.
By contrast, our model does not make this assumption.


\subsection{Information Sharing for MARL}
For various cooperative tasks where explicit coordination is required to solve the task, enabling communication to share information across the agents helps with the performance.
In CommNet \cite{CommNet}, the authors introduce a model to learn a differentiable communication protocol between multiple agents. However, it does not explicitly model the interactions between the agents but rather averages the states of all the neighboring agents. VAIN \cite{VAIN} improves upon CommNet by using an exponential kernel-based attention to choose specific messages from other agents to attend to. Similarly, ATOC \cite{ATOC} and TarMAC \cite{TarMAC} use an attention mechanism for communication among agents but without any restrictions on which agent can communicate with which others, leading to centralization during execution. 
GAXNET \cite{GAXNet} also uses an attention mechanism \cite{AttentionIsAllYouNeed}, but additionally allows for the exchanging of weights with other agents to reduce the attention mismatch between them. While their model is shown to work well for navigation, it requires the maximum number of agents in the environment to be fixed before training. We also use an attention mechanism for inter-agent communication. However, we do not have any message passing between objects in the environment (also called 'entities') and agents; instead, the agents themselves do all the computation with local information of the entities’ states.

%% file: sections/3_methodology.tex
    \begin{figure*}[t!]
        \centering
        \includegraphics[width=0.92\textwidth]{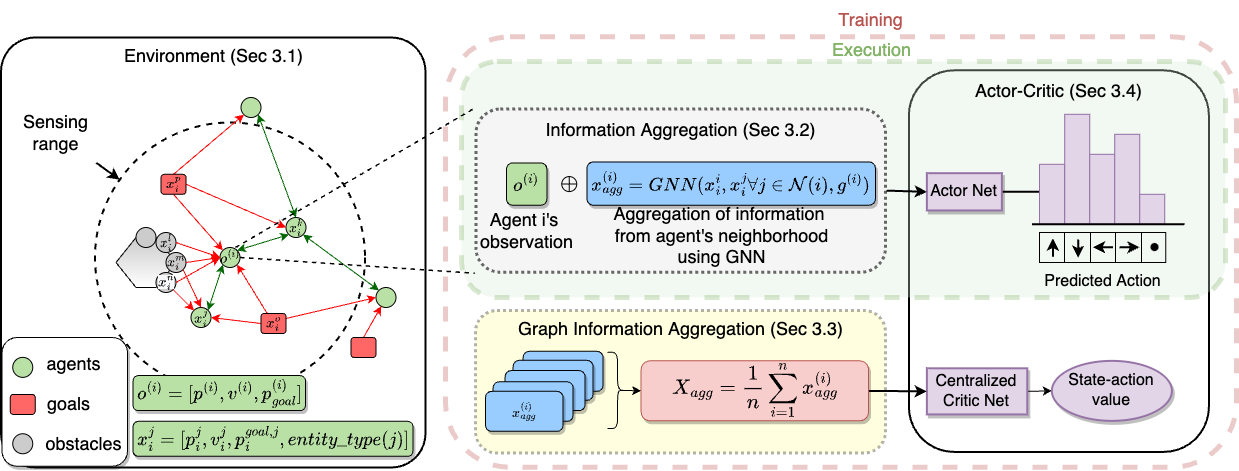}
        \caption{Overview of InforMARL. (i) Environment: The agents are depicted by green circles, the goals are depicted by red rectangles, and the unknown obstacles are depicted by gray circles. $x^{(i)}_{agg}$ represents the aggregated information from the neighborhood, which is the output of the GNN. A graph is created by connecting entities within the sensing-radius of the agents. (ii)  Information Aggregation: Each agent's observation is concatenated with $x^{(i)}_{\mathrm{agg}}$. The inter-agent edges are bidirectional, while the edges between agents and non-agent entities are unidirectional. (iii) Graph Information Aggregation: The aggregated vector from all the agents is averaged to get $X_{\mathrm{agg}}$.
        (iv) Actor-Critic: The concatenated vector $[o^{(i)}, x^{(i)}_{\mathrm{agg}}]$ is fed into the actor network to get the action, and $X_{\mathrm{agg}}$ is fed into the critic network to get the state-action values.}
        \label{fig:arch}
    \end{figure*}
    
\section{Description of InforMARL}
    \label{section:method}
    Our MARL framework for navigation, InforMARL, consists of four modules, as shown in Figure \ref{fig:arch} and described below. 
    \subsection{Environment}
        \label{section:env}
        Every object in the environment (also known as an `entity') is assumed to be either an agent, an obstacle, or a goal/landmark. We define an agent-entity graph with respect to agent $i$ at each time-step $t$, as $g^{(i)}_t \in \mathcal{G}: (\mathcal{V}, \mathcal{E})$, where each node $v\in\mathcal{V}$ is an entity in the environment. The variable \texttt{entity\_type(j)}$\in\{ \texttt{agent}, \texttt{obstacle}, \texttt{goal}\}$ determines the type of entity at node $j$. 
        There exists an edge $e\in\mathcal{E}$ between an agent and an entity if they are within a `sensing radius' $\rho$ of each other. The agent-agent edges are bidirectional, whereas the agent-non-agent edges are unidirectional (\ie messages can only be passed from the non-agent entity to the agent). In other words, a unidirectional edge is equivalent to the agent sensing a nearby entity's state, while a bidirectional edge is equivalent to a communication channel  between agents. This structure is similar to the agent-entity graph defined in \cite{EMP}, but without the assumption that all agents have access to the positions of all entities at the beginning of each episode. Note that our formulation also supports cases where disconnected sub-graphs are formed due to the positioning of the entities in the environment. 

        The corresponding Decentralized Partially Observable Markov Decision Process (Dec-POMDP) \cite{MDP, POMDP1, POMDP2} is characterized by the tuple $\langle N, S, O, \mathcal{A}, \mathcal{G}, P, R,\gamma \rangle$, where $N$ is the number of agents, $s\in S = \mathbb{R}^{N\times D}$ is the state space of the environment and $D$ is the dimension of the state, and $o^{(i)}=O(s^{(i)})\in \mathbb{R}^{d}$ is the local observation for agent $i$, where $d\leq D$ is the observation dimension. 
        $a^{(i)} \in \mathcal{A}$ is the action space for agent $i$ and the joint action for all $N$ agents is given by $A=(a^{(1)},\cdots,a^{(N)})$. 
        Specifically, $a^{(i)}$ is a one-hot vector of size equal to the number of possible actions. $g^{(i)} \in \mathcal{G}(s; i)$ is the graph network formed by the entities in the environment with respect to agent $i$. $P(s'|s, A)$ is the transition probability from $s$ to $s'$ given the joint action $A$. $R(s, A)$ is the joint reward function. $\gamma \in [0,1)$ is the discount factor. The MARL training process seeks to find an optimal policy, $\Pi= \left(\pi^{(1)}, \cdots, \pi^{(N)}\right)$, where each agent uses a policy $\pi_\theta^{(i)}\left(a^{(i)}|o^{(i)}, g^{(i)}\right)$ parameterized by $\theta$ to determine its action $a^{(i)}$ from its local observation $o^{(i)}$ and the graph network $g^{(i)}$ that it is a part of, while optimizing the discounted accumulated reward $J(\theta) = \mathbb{E}_{A_t, s_t}\left[\sum\limits_t \gamma^t R\left(s_t, A_t\right)\right]$.
    
        Agent $i$'s local observation $o^{(i)}$ consists of its position and velocity in a global frame of reference and the relative position of the agent's goal with respect to its position.
        Each node $j$ on the graph $g^{(i)}$ has node features $x_j=[p^j_i, v^j_i, p^{\mathrm{goal},j}_i, \texttt{entity\_type(j)}]$ where $p^j_i, v^j_i, p^{\mathrm{goal},j}_i$ are the \emph{relative} position, velocity, and position of the goal of the entity at node $j$ with respect to agent $i$, respectively. If node $j$ corresponds to a (static/dynamic) obstacle or a goal, we set $p^{\mathrm{goal},j}_i \equiv p^j_i$. 
         Each edge $e_{ij}$ has an associated edge feature given by the Euclidean distance between the entities $i$ and $j$. For processing the \texttt{entity\_type} categorical variable, we experimented with both using an embedding layer \cite{word2vec} and using one-hot encoding. No significant performance advantage with one method over the other was found, so we chose to use the embedding layer. Further analysis revealed that the learned embedding vectors for \texttt{entity\_type} were equidistant from each other when visualized in 2D. This was to be expected because each of the entity types $(agent, obstacle, goal)$ are equally distinct from another. Future work could include further refinement of \texttt{entity\_type}, \eg, adversarial vs. cooperative obstacles, static vs. dynamic obstacles, etc.

We adopt a similar reward function as used in multi-agent particle environment (MAPE) 
 \cite{MAPE} where the joint reward function is defined as $R(s_t,A_t)=\sum_{i=1}^{N}r_t^{(i)}$, which encourages cooperation among all agents. Here $r_t^{(i)}$ is each agent's reward at timestep $t$ and depends on the scenario. Details about the reward functions for the different task environments can be found in Appendix \ref{appendix:env}.

\subsection{Information Aggregation}
    \label{section:info_agg}
    To infer information about the local neighborhood around each agent $i$, we use a graph neural network (GNN) with a message passing framework \cite{message_passing_GNN}. Specifically, we use Unified Message Passing Model (UniMP) \cite{graphTransformerConv}, a variant of a graph transformer \cite{AttentionIsAllYouNeed, graphTransformer} where each layer update is defined as $x_i' = W_1 \cdot x_i + \sum\limits_{j\in\mathcal{N}(i)}\alpha_{i,j}W_2 \cdot x_j$, 
    where $x_k$ are the node features in the graph, $\mathcal{N}(i)$ is the set of nodes which are connected to node $i$, $W_k$ are learnable weight matrices and the attention coefficients $\alpha_{i,j}$ are computed via multi-head dot product attention:
    \begin{equation}
        \alpha_{i,j} = \mathrm{softmax}\left(\dfrac{\left(W_3 \cdot x_i\right)^T\left(W_4 \cdot x_j + W_5 \cdot e_{ij}\right)}{\sqrt{\mathrm{c}}}\right)
    \end{equation}
    where 
    $e_{ij}$ are edge features for the edge connecting nodes $i$ and $j$, and $c$ is the output dimension for that specific layer.
    
    The attention mechanism allows the agents to selectively prioritize messages coming from their neighbors according to their importance. We use multiple layers of this message-passing so that information can be propagated between agents that are higher-order neighbors with each other. For each agent $i$, this module aggregates information from the neighboring nodes in the graph into a fixed-sized vector $x^{(i)}_{\mathrm{agg}}$. The concatenated vector $[o^{(i)}, x^{(i)}_{\mathrm{agg}}]$ is given as the input to the actor network. This architecture allows InforMARL to dynamically adapt to a changing number of entities in the environment while remaining invariant to the permutation of the observed entities.

\subsection{Graph Information Aggregation}
    \label{section:graph_info_agg}
    While training a model in the CTDE setting, the critic generally gets the state-action pairs of all individual agents in the environment as a concatenated vector. To make the training transferable to a variable number of agents and to aid with curriculum learning \cite{curriculum_learning, transfer_learning1, transfer_learning2}, we replace this concatenation with a graph information aggregation module. This module is similar to the `Information Aggregation' one in which a GNN aggregates information from the agent's neighbors. A global mean pooling operator,  $X_{\mathrm{agg}} = \frac{1}{N}\sum\limits_{i=1}^N x^{(i)}_{\mathrm{agg}}$, is applied to aggregate the updated node features in the graph. 
    Note that $X_{\mathrm{agg}}$ is a vector of fixed size independent of the number of agents, which is not the case when concatenating the state action-pairs of all individual agents. This vector is then given as input to the critic network.

\subsection{Actor-Critic Networks}
    \label{section:actor_critic}
    The actor and critic networks can be either a multi-layer perceptron (MLP) or a recurrent neural network (RNN) \cite{rnnForPOMDP}, using either LSTMs \cite{LSTM} or GRUs \cite{GRU}. Our proposed information aggregation method can be used in conjunction with any standard MARL algorithm (\eg, MADDPG \cite{MADDPG}, MATD3 \cite{MATD3}, MAPPO \cite{MAPPO}, QMIX \cite{QMIX}, VDN \cite{VDN}, etc.).

%% file: sections/4_experiments.tex
\section{Experiments}
\label{section:experiments}



    

\textbf{Environment descriptions}: We evaluate our proposed model on four different navigation tasks by modifying the MAPE \cite{MAPE}. In all these environments, $N$ agents move around in a 2D space following a double integrator dynamics model \cite{double_integrator_model}. Each agent has a discrete action space where it can control unit acceleration and deceleration in the $x$- and $y$- directions. 
\begin{CompactEnumerate}
    \item \textbf{Target}: 
    Each agent tries to reach its preassigned goal while avoiding collisions with other entities in the environment.
    \item \textbf{Coverage} \cite{targetCoverage1, targetCoverage2}: Each agent tries to go to a goal while avoiding collisions with other entities, and ensuring that no more than one agent reaches the same goal.
    \item \textbf{Formation} \cite{EMP}: There is a single landmark (the counterpart of a goal for this task), and the agents try to position themselves in an $N$-sided regular polygon with the landmark at its centre.
    \item \textbf{Line} \cite{EMP}: There are two landmarks, and the agents try to position themselves equally spread out in a line between the two.
\end{CompactEnumerate}
We first focus on the \emph{Target} task environment in this section, and present the performance of InforMARL on the other tasks in Section \ref{sec:other_tasks}. More details about the environments can be found in Appendix \ref{appendix:env}.

\noindent \textbf{Implementation specifications}:
    We chose to use MAPPO \cite{MAPPO} as the base MARL algorithm for InforMARL because it was found to be the best performing of the standard MARL baselines in the 3 agent-3 obstacle \emph{Target} environment. We implemented InforMARL by modifying the official codebase for MAPPO in PyTorch. The codebase links to our baseline implementations can be found in Appendix \ref{appendix:implementation}. 
    We used the official implementations for most of the baselines considered in Section \ref{sec:comparison_algos}. We prefix the algorithm name with `R' to denote the recurrent neural network (RNN) version of the algorithm (\eg MAPPO-RMAPPO, \etc).
    We use the same hyperparameters as used in the experiments for MAPPO, and do not perform parameter tuning on InforMARL for any of the MAPPO-based parameters. The hyperparameters used in our experiments can be found in Appendix \ref{appendix:hyperparams}.

\noindent \textbf{Amount of information available to agents}: The amount of information available to each agent determines whether or not it can learn a meaningful policy. Although having more information generally translates to better performance, it does not necessarily scale well with the number of agents. Prior works \cite{MAPPO, MADDPG} have typically used a na{\"i}ve concatenation of the states of all agents or entities in the environment fed into a neural network. Such an approach scales poorly (the network input size is determined by the number of agents) and does not transfer well to scenarios with a different number of agents than the training environment. We vary the amount of information available to agents by defining three information modes: 
    \begin{itemize}
        \itemsep0em 
        \item \emph{Local:} In the local information mode, $o^{(i)}_{\mathrm{loc}} = [p^{(i)}, v^{(i)}, p^{(i)}_{\mathrm{goal}}]$ where $p^{(i)}$ and $v^{(i)}$ are the position and velocity of agent $i$ in a global frame, and $p^{(i)}_{\mathrm{goal}}$ is the position of the goal relative to the agent's position. InforMARL uses this information mode.
        \item \emph{Global:} Here, $o_{\mathrm{glob}}^{(i)} = [p^{(i)}, v^{(i)}, p^{(i)}_{\mathrm{goal}}, p^{(i)}_{\mathrm{other}}]$, where $p^{(i)}_{\mathrm{other}}$ comprises of the relative positions of all the other entities in the environment. The scenarios defined in the MAPE (and consequently, other approaches that use MAPE) use this type of information mode unless explicitly stated otherwise.
        \item \emph{Neighborhood:} Here, agent $i$ observes $o_{\mathrm{nbd}}^{(i)} = [p^{(i)}, v^{(i)}, p^{(i)}_{\mathrm{goal}}, p^{(i)}_{\mathrm{other}}]$, where $p^{(i)}_{\mathrm{other}}$ comprises of the relative positions of all other entities which are within a distance \texttt{nbd-dist} of the agent. The maximum number of entities within the neighborhood is denoted \texttt{max-nbd-entities}, and so the dimension of the observation vector is fixed. If there are fewer than \texttt{max-nbd-entities} within a distance \texttt{nbd-dist} of the agent, we pad this vector with zeros.
    \end{itemize}
    
\subsection{A Motivating Experiment \label{sec:motivating_expt}}
\begin{figure}[h!]
        \centering
        \includegraphics[width=0.8\linewidth]{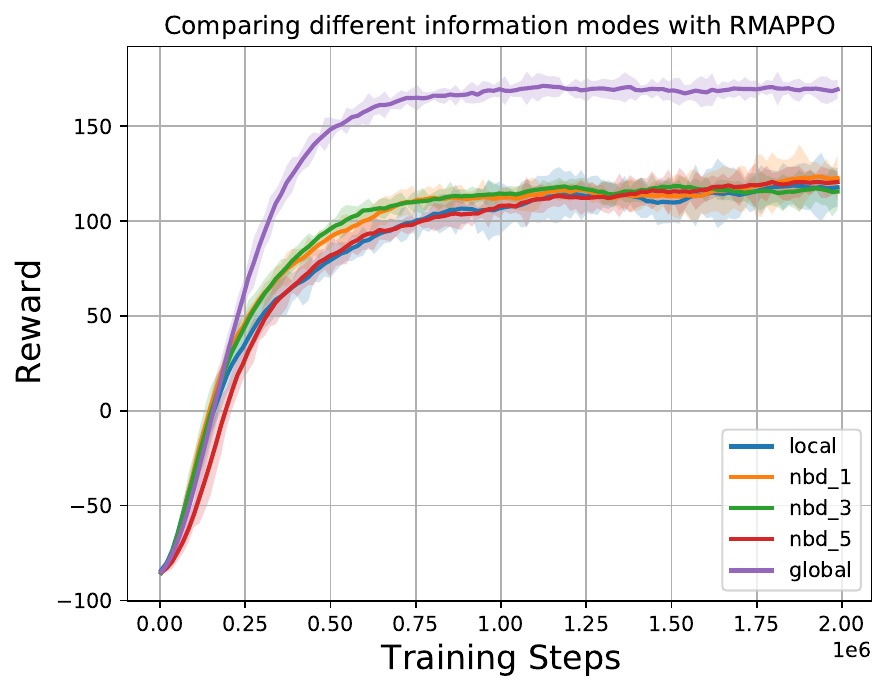}
         \setlength{\belowcaptionskip}{-12pt}
        \caption{RMAPPO with the local, neighborhood (with 1, 3, and 5 \texttt{max-nbd-entities}), and global information given as states. The plots show the rewards during training for the 3 agent-3 obstacle \emph{Target} task environment. 
        Comparing the global information mode to the others, we see that merely providing local information and a na{\"i}ve concatenation of neighborhood information is not sufficient to learn an optimal policy.}
        \label{fig:local_nbd_global}
    \end{figure}

    The \emph{local} or \emph{neighborhood} information modes are transferable to scenarios with a different number of entities in the environment. By contrast, the \emph{global} information mode is not transferable to other scenarios. Figure \ref{fig:local_nbd_global} shows the rewards obtained during training using the three information modes in the \emph{Target} task environment. 
    
   We see in Figure \ref{fig:local_nbd_global}  that the policy learned with global information is better than the policies learned with local or neighborhood information. This is because it has the information necessary to take optimal actions. In the `nbd\_5' scenario, if all entities are within the \texttt{nbd\_dist} ball, then $o^{(i)}_{\mathrm{nbd}}\equiv o^{(i)}_{\mathrm{glob}}$, since there are only five entities in the environment apart from the agent itself.
    Although the `nbd\_5' scenario is almost similar to the global information mode, the performances are not similar: the global information mode achieves much higher rewards. This behavior is because the observation $o_{i,\mathrm{nbd}}$ can change temporally from having information about an entity when it lies within a \texttt{nbd\_dist} ball of the agent, and then getting padded with zeros when it is out of the ball. This inconsistency in the amount of information accessible to the agent at every time step causes a significant performance difference between the policies learned with global and neighborhood information modes. This experiment motivates us to find a way in which more information can be leveraged (similar to the global case), but in a manner that does not suffer from the performance shortcomings of the neighborhood or local information modes.

\subsection{Comparison of InforMARL with Other Baselines} \label{sec:comparison_algos}
 \begin{figure*}[h!]
\includegraphics[width=0.33\textwidth]{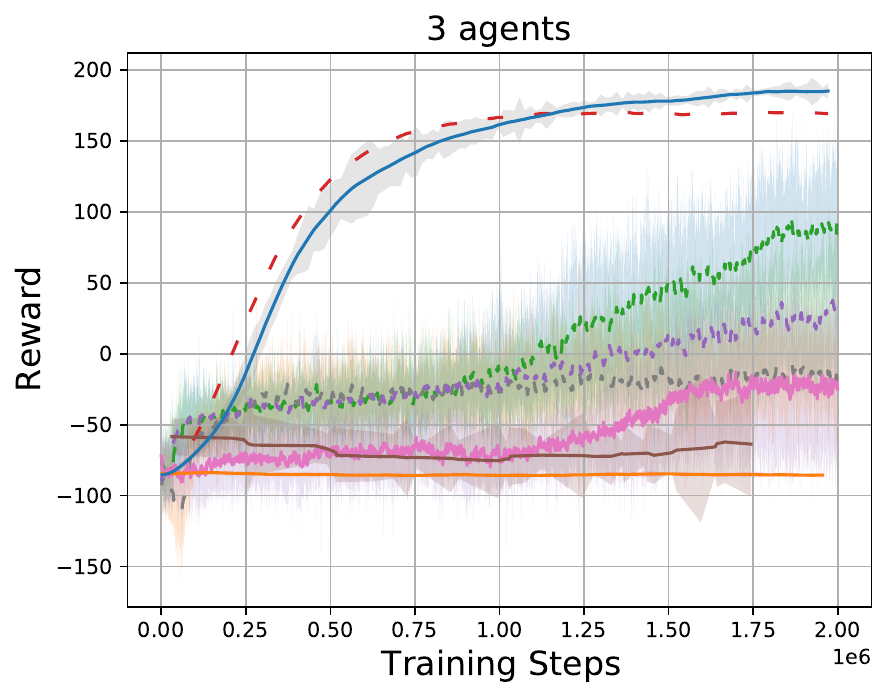}
  %
            \includegraphics[width=0.33\textwidth]{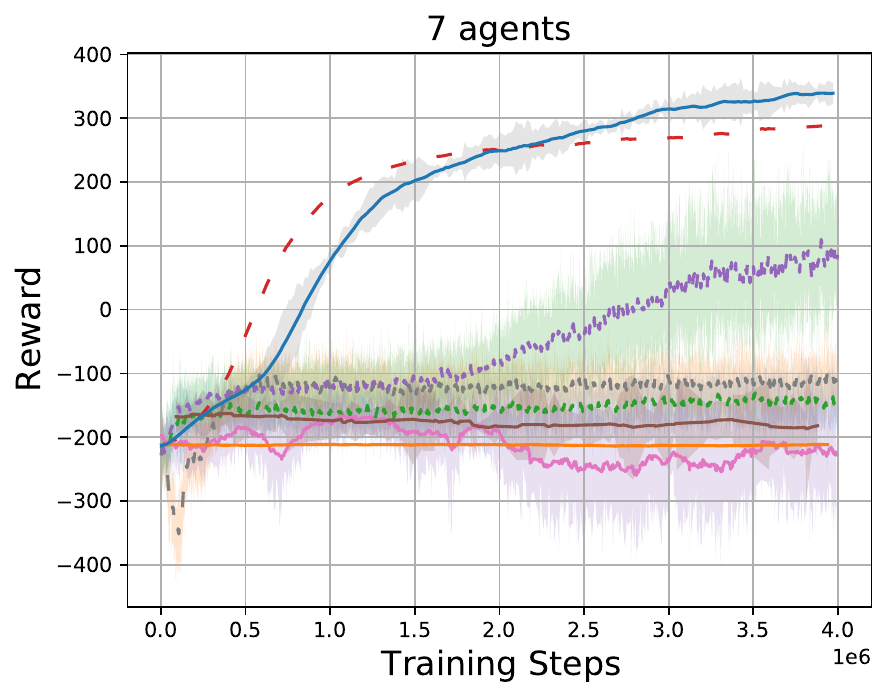}
 %
  \includegraphics[width=0.33\textwidth]{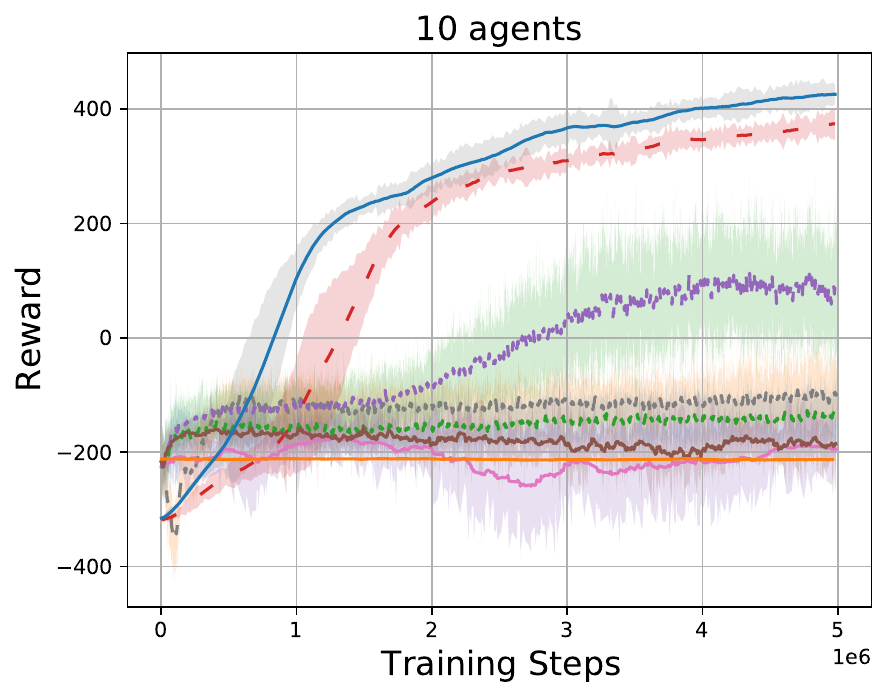}
 %
      \centering   \includegraphics[width=0.8\textwidth]{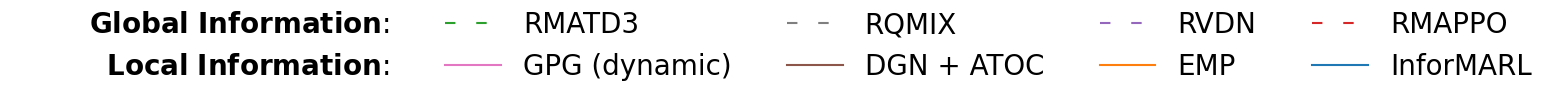}
        \caption{Comparison of the training performance of InforMARL with the best-performing baselines using global and local information in the \emph{Target} task environment. The means and standard deviations of the rewards over training with five random seeds are shown. InforMARL significantly outperforms most baseline algorithms. Although RMAPPO has similar performance, it requires global information. Appendix \ref{appendix:full_comparison} presents a complete comparison to more baselines.}
        \label{fig:compare_plots}
    \end{figure*}
    
    As shown in Section \ref{sec:motivating_expt}, using local information modes or na{\"i}vely concatenating neighborhood entity information is not sufficient to learn optimal policies. In this section, we demonstrate that InforMARL can effectively learn policies for navigation given local information. We then compare its performance  in the \emph{Target} environment with prior deep MARL approaches. Specifically, we consider the following methods as baselines for comparison.
    \begin{CompactEnumerate}
   \item \textbf{Graph Policy Gradient (GPG) \cite{GPG}:} GPG uses a graph convolutional neural network (GCN) \cite{GCN} to parameterize policies for agents. The authors use the policy gradient method \cite{policyGradient} as the base MARL algorithm. We perform experiments with both dynamic and static graphs. \textit{(Note: It was shown in \cite{GPG} that using a static graph constructed at the beginning of the episode was better than using a dynamic graph.)}
    \item \textbf{Graph Convolutional Reinforcement Learning (DGN, DGN+ATOC) \cite{ATOC, DGN}:} Similar to GPG, these methods use GCNs to capture interactions between the agents in the environment. A key difference between InforMARL and these two methods (GPG and DGN) is that while the latter two approaches consider only agents in their message-passing graph, InforMARL also includes other (non-agent) entities in the graph.
    \item \textbf{Entity Message Passing (EMP) \cite{EMP}:} Similar to InforMARL, EMP uses an agent-entity graph. However, in contrast to InforMARL, EMP assumes that agents know the positions of all entities in the graphs (i.e., global information) at the beginning of the episode. 
    \item \textbf{Other standard MARL Algorithms:} Finally, we compare InforMARL with standard MARL algorithms, namely, MADDPG \cite{MADDPG}, MATD3 \cite{MATD3}, QMIX \cite{QMIX}, VDN \cite{VDN} and MAPPO \cite{MAPPO}. In each case, we also consider the recurrent neural network versions. We focus on results for the global information modes, we found that these methods did not learn well with just local information.
    \end{CompactEnumerate}
    
    Figure \ref{fig:compare_plots} shows the training performance of InforMARL and the  best-performing of the baselines mentioned above for the \emph{target} task environment. For ease of visualization, we plot only the four best-performing methods with global and local information, respectively. Each line corresponds to the mean and standard deviation over five random seeds.  We consider scenarios with $N = \{3, 7, 10\}$ agents. Comparisons to other baselines can be found in Appendix \ref{appendix:full_comparison}, and ablation studies for varying sensing radii in Appendix \ref{appendix:ablation}.
    
   Figure \ref{fig:compare_plots} illustrates that only using RMAPPO (\ie the RNN version of MAPPO) or InforMARL, agents are able to learn to navigate and get to their goals. However, unlike RMAPPO which requires global information, InforMARL achieves this with just local information. Furthermore, InforMARL requires a similar number of training steps as RMAPPO, despite having access to much less information.    
    
    \begin{table*}\small
        \begin{tabular}{|c||p{0.3in}||c|c|r|c||c|c|r|c||c|c|r|c|}
        \hline
        \multirow{2}{*}{\bf Algorithm} & {\bf Info} & \multicolumn{4}{c||}{$\boldsymbol{N=3}$} & \multicolumn{4}{c||}{$\boldsymbol{N=7}$} & \multicolumn{4}{c|}{$\boldsymbol{N=10}$}\\
        \cline{3-14}
         & {\bf mode} & Reward & $T$ & \# col  & $S\%$ & Reward & $T$ & \# col & $S\%$ & Reward & $T$ & \# col & $S\%$\\
    \hline \hline
        RMADDPG          & Global          &   $10.73$     &  $0.75$ &         $1.72$   &  23 & $-122.07$  & $0.95$  & $4.61$ &  3 &     -127.98          &  1.00      &  7.39 &      0         \\ \hline
        RMATD3           & Global          &   $105.49$     & $0.51$  &     $1.62$      &  67  &     $-128.93$   & $0.96$  &      $3.67$         &  6  &  -131.72  & 0.99   &  5.94  &    1       \\ \hline
        RQMIX           & Global          &    $19.21$    &  $0.77$ &       $0.83$     & 28  &   $-83.41$     & $0.85$  &    $5.83$      & 12 &   -76.98     &  0.96 &  8.65  &     2      \\ \hline
        RVDN             & Global          &    $64.04$    &  $0.62$ &          $0.57$     &  45  &  $140.94$  &  $0.62$ &       $3.42$   &   47  &    157.63    &  0.64 &  4.93  &  43           \\ 
        \specialrule{.15em}{.0em}{0em} 
            \textbf{RMAPPO}           & \textbf{Global}          &   \textbf{173.13}     & \textbf{0.41}  &        \textbf{0.67}   &  \textbf{96} &     \textbf{327.39}   & \textbf{0.44}  &      \textbf{4.29}    &  \textbf{88}    &  \textbf{366.81}      &  \textbf{0.44} &   \textbf{6.93}  &  \textbf{79}        \\ \specialrule{.15em}{.0em}{0em} 
        GPG (dyn.)    & Local       &   $-46.27$     &  $0.87$ &     $0.21$     &   8  &  $-165.91$       & $1.00$  &        $1.57$       &  3  & -173.53  &  1.00 &   2.24   &    0     \\ \hline
        DGN+ATOC       & Local       &    $67.70$    &  $0.66$ &      $0.72$         &  35  &  $-189.61$  & $0.97$  &      $1.03$    &  0   &   -201.01     & 1.00  &   1.97   &  0       \\ \hline
        EMP              & Local       &  $-83.96$      &  $0.98$ &       $0.72$     & 6   &  $-211.90$      & $0.98$  &     $2.63$   &    0  & -209.90   & 1.00  &  94.12   &    0      \\ 
        \specialrule{.15em}{.0em}{0em} 
        \textbf{InforMARL} & \textbf{Local} &     \textbf{205.24}   & \textbf{0.17}  &       \textbf{1.45}    &  \textbf{100}  &     \textbf{399.01}   &  \textbf{0.37} &  \textbf{3.72}    &  \textbf{100}       &  \textbf{429.14}      &  \textbf{0.39} &  \textbf{4.73}   & \textbf{100}          \\ 
         \specialrule{.15em}{.0em}{0em} 
        \end{tabular}
        \setlength{\belowcaptionskip}{-12pt}
        \caption{Comparison of InforMARL with the best-performing baseline methods, for the \emph{Target} task environment with 3, 7, and 10 agents, averaged over 100 test episodes. See Appendix \ref{appendix:full_comparison} for comparisons to other baseline methods.
        }
        \label{table:comparison}
    \end{table*}
    
    We present the following metrics in Table \ref{table:comparison}. The results represent an average over 100 test episodes.
    \begin{CompactEnumerate}
        \item The total rewards obtained by the agents during an episode 
        A higher value corresponds to better performance.
        \item The fraction of an episode that the agents take on average to get to the goal, denoted $T$. If an agent does not reach its goal, then $T$ is set to be 1 (lower is better).
        \item Percent of episodes in which all agents are able to get to their goals, denoted $S$\% (higher is better).
        \item The total number of collisions (both agent-agent + agent-obstacle) that agents had in an episode, denoted \# col. The lower this metric, the better the performance of the algorithm. 
    \end{CompactEnumerate}
    Although having a smaller number of collisions is better, the policies of some of the baseline algorithms do not significantly move the agents from their initial position after training and hence do not get to the goal. This leads to them having a lower number of collisions. Hence, this metric should be judged with the success rate in context.
    
    The graph-based methods, namely GPG (static and dynamic), DGN (+ATOC), and EMP do not learn effectively with local information modes. Although both GPG and DGN use GCNs, they do not perform as well as InforMARL because they use only agent-agent and not agent-entity graphs. The lack of information about non-agent entities means that agents cannot maneuver through the environment to avoid collisions.
   
   In contrast to the results showed in \cite{GPG}, our results show that the dynamic graph version of GPG is slightly better than the static graph one. A possible reason for this discrepancy is that the fixed formation environment considered in \cite{GPG} is more amenable to the use of static graphs than the navigation environment. EMP, which uses a similar agent-entity graph as ours, fails to learn because of the strong assumption of having access to the positions of all entities in the environment at the beginning of the episode. The original implementation of EMP (and the associated environments) included information about all the entities other than agents in the observation vector.

\subsection{Scalability of InforMARL}
    To evaluate the scalability of InforMARL with minimum information, we perform experiments in the \emph{Target} environment by testing the models in scenarios with a different number of agents from those they were trained on. 

    \begin{table}[h]\small\centering
        \begin{tabular}{|l|c||c|c|c|}
        \hline
        \multicolumn{2}{|c||}{\diagbox{\bf Test}{\bf Train}}& $\boldsymbol{n=3}$  & $\boldsymbol{n=7}$ & $\boldsymbol{n=10}$\\ 
        \hline\hline
            \multirow{4}{*}{$\boldsymbol{m=3}$} 
            & Reward/$m$ & 63.21 & 63.25 & 62.87\\ \cline{2-5}
            & $T$ & 0.39 & 0.40 & 0.40\\ \cline{2-5}
            & (\# col)/m & 0.40 & 0.46 & 0.49\\ \cline{2-5}
            & $S\%$ & 100 & 100 & 99\\
             \hline\hline
            
            \multirow{4}{*}{$\boldsymbol{m=7}$} 
            & Reward/$m$ & 61.16 & 62.23 & 61.32\\ \cline{2-5}
            & $T$ & 0.38 & 0.40 & 0.40\\ \cline{2-5}
            & (\# col)/$m$ & 0.74 & 0.66 & 0.70\\ \cline{2-5}
            & $S\%$ & 100 & 100 & 100\\
              \hline\hline
            
            \multirow{4}{*}{$\boldsymbol{m=10}$} 
            & Reward/$m$ & 58.59 & 58.23 & 58.67\\ \cline{2-5}
            & $T$ & 0.38 & 0.40 & 0.39\\ \cline{2-5}
            & (\# col)/$m$ & 0.95 & 0.88 & 0.87\\ \cline{2-5}
            & $S\%$ & 100 & 99 & 100\\ 
            \hline \hline 
            \multirow{4}{*}{$\boldsymbol{m=15}$} 
            & Reward/$m$ & 53.19 & 53.46 & 54.21\\ \cline{2-5}
            & $T$ & 0.39 & 0.40 & 0.40\\ \cline{2-5}
            & (\# col)/$m$ & 1.28 & 1.21 & 1.20\\ \cline{2-5}
            & $S\%$ & 100 & 99 & 99\\
            \hline
        \end{tabular}
        \caption{Test performance of InforMARL for the \emph{Target} task, when trained on scenarios with $n$ agents and tested with $m$ agents in the environment.
        }
        \label{table:transfer}
    \end{table}
    
    Table \ref{table:transfer} shows the results of testing InforMARL trained on $n$ agents and tested on $m$ agents. Each scenario is tested over 100 episodes. The number of obstacles in the environment is randomly chosen from $(0,10)$ at the beginning of the episode. Based on the findings presented in Table \ref{table:comparison}, we did not test the scalability of other methods since they did not perform well in the local information mode, or could not handle varying numbers of entities in the environment.  
    
    We see from  Table \ref{table:transfer} that InforMARL is able to achieve a success rate of 100\% for almost all the scenarios when the number of agents in the environment as varied. Furthermore, with InforMARL, the agents are able to get to their goals within $T \sim 0.39$ of the episode length in all scenarios. The number of collisions per agent increases, which is to be expected as the environment becomes denser. We believe that this can be remedied by using a stricter penalty 
    for collisions. 
    Alternatively, 
    control barrier functions for satisfying safety constraints could be used to provide a formal safety guarantee in the MARL setting \cite{MARL_control_barrier}.

\subsection{Performance in Other Task Environments\label{sec:other_tasks}}
In the \emph{Target} task environment, coordination amongst agents was required only for collision avoidance (both with other agents and obstacles) since the goal positions for all the agents were predetermined. For the \emph{Coverage}, \emph{Formation} and \emph{Line} tasks, the agents not only need to coordinate for collision avoidance but also need to develop  consensus on their goals. Table \ref{table:transfer2} shows the success rate and the fraction of episode taken to complete the task in the \emph{Coverage}, \emph{Formation} and \emph{Line} environments. Here, InforMARL was trained in the 3-agent scenario and tested on the 3- and 7-agent scenarios, whereas RMAPPO is trained and tested on scenarios with the same number of agents. InforMARL is able to achieve a success rate of almost 100\% across all scenarios in the different environments, while taking a similar fraction of the episode to complete as RMAPPO. While RMAPPO requires global information to learn a successful policy, InforMARL only needs local neighborhood information. This illustrates the effectiveness of the information aggregation module in InforMARL when the agents only have access to local information.



    \begin{table}[h]
    \small\centering
        \begin{tabular}{|c|c|c|c|c|}
        \hline
        \multirow{2}{*}{\bf Environment}& \multirow{2}{*}{$\boldsymbol{m}$} & \multirow{2}{*}{\bf Metric} & \multicolumn{2}{c|}{\bf Algorithm}\\ 
        \cline{4-5}
        & & &  {\bf RMAPPO} & {\bf InforMARL}\\
        \hline\hline
        \multirow{4}{*}{\bf Coverage} & \multirow{2}{*}{3}& $T$ &  0.34& 0.36 
        \\  \cline{3-5}
        & & $S\%$ &  100 & 100
        \\  \cline{2-5}
        & \multirow{2}{*}{7}& $T$ &   0.42& 0.43 
        \\  \cline{3-5}
        & & $S\%$ &  100 & 99
        \\   \specialrule{.12em}{.0em}{0em}
        \multirow{4}{*}{\bf Formation} & \multirow{2}{*}{3}& $T$ & 0.31 & 0.30 
        \\  \cline{3-5}
        & & $S\%$ & 100 & 100
        \\  \cline{2-5}
        & \multirow{2}{*}{7}& $T$ &  0.47& 0.43
        \\  \cline{3-5}
        & & $S\%$ & 100& 100
        \\   \specialrule{.12em}{.0em}{0em}
        \multirow{4}{*}{\bf Line} & \multirow{2}{*}{3}& $T$ &  0.24& 0.21 \\  
        \cline{3-5}
        & & $S\%$ &  100 & 100
        \\  \cline{2-5}
        & \multirow{2}{*}{7}& $T$ & 0.38& 0.36  \\
        \cline{3-5}
        & & $S\%$ & 100 & 100
        \\  \hline
        \end{tabular}
         \setlength{\belowcaptionskip}{-8pt}
        \caption{Performance of RMAPPO and InforMARL on the \emph{coverage}, \emph{formation}, and \emph{line} tasks. We note that InforMARL was trained on the 3-agent scenario and tested on $m=\{3,7\}$ agents, while RMAPPO was trained and tested on the same number of agents (\ie with $m=n$). 
        }
        \label{table:transfer2}
    \end{table}

%% file: sections/5_conclusion_future.tex
\section{Conclusions and Future Work}
\label{section:conclusion_future_work}
We introduced InforMARL, a novel architecture that uses GNNs for scalable multi-agent reinforcement learning. We showed that having just local observations as states is not enough for standard MARL algorithms to learn meaningful policies. Along with this, we also showed that albeit na{\"i}vely concatenating state information about all the entities in the environment helps to learn good policies, they are not transferable to other scenarios with a different number of entities than what it was trained on. InforMARL is able to learn transferable policies using standard MARL algorithms using just local observations and an aggregated neighborhood information vector. Furthermore, it has better sample complexity than other standard MARL algorithms that use global observation. We demonstrated these findings for four environments with different navigation tasks: \emph{target}, \emph{coverage}, \emph{formation}, and \emph{line}.  
Future work will include the introduction of more complex (potentially adversarial) dynamic obstacles in the environment and adding a safety guarantee layer for the actions of the agents to avoid collisions. Additionally, the use of InforMARL for curriculum learning and transfer learning to different environments 
is a topic of ongoing research. 

%% file: sections/6_acknowledgement.tex
\section*{Acknowledgements}
The authors would like to thank the MIT SuperCloud
\cite{supercloud} and the Lincoln Laboratory Supercomputing Center for providing high performance computing resources that have contributed to the research results reported
within this paper. The NASA University Leadership initiative (grant \#80NSSC20M0163) provided funds to assist the authors with their research, but this article solely reflects the opinions and conclusions of its authors and not any NASA entity. This research was sponsored in part by the United States AFRL and the United States Air Force Artificial Intelligence Accelerator and was accomplished under Cooperative Agreement Number FA8750-19-2-1000. The views and conclusions contained in this document are those of the authors and should not be interpreted as representing the official policies, either expressed or implied, of the United States Air Force or the U.S. Government. The U.S. Government is authorized to reproduce and distribute reprints for Government purposes notwithstanding any copyright notion herein.

%% file: sections/7_appendix.tex
\appendix
\section{Baseline Implementation Sources}
\label{appendix:implementation}
We modified the codebases from the official implementations for the GPG, DGN, EMP, and MAPPO baselines and a thoroughly benchmarked codebase for MADDPG, MATD3, VDN and QMIX and provide the links to those implementations here. Note that we used the same hyperparameters as used in their original implementations assuming that they were optimal. We performed a hyperparameter search for these algorithms by varying the learning-rates, network size and a few algorithm specific parameters but did not find a better set of hyperparameters for the environment and chose to report with the original hyperparameters. 
\begin{itemize}
    \item GPG: \href{https://github.com/arbaazkhan2/gpg_labeled}{https://github.com/arbaazkhan2/gpg\_labeled}
    \item DGN: \href{https://github.com/jiechuanjiang/pytorch_DGN}{https://github.com/jiechuanjiang/pytorch\_DGN}
    \item EMP: \href{https://github.com/sumitsk/marl_transfer}{https://github.com/sumitsk/marl\_transfer}
    \item MAPPO: \href{https://github.com/marlbenchmark/on-policy}{https://github.com/marlbenchmark/on-policy}
    \item MADDPG, MATD3, QMIX, VDN: 
    \href{https://github.com/marlbenchmark/off-policy}{https://github.com/marlbenchmark/off-policy}
\end{itemize}

\section{Environment Tasks}
\label{appendix:env}
\begin{figure*}[htp]
  \centering
  \subfigure[Target]{\fbox{\includegraphics[height=0.22\linewidth]{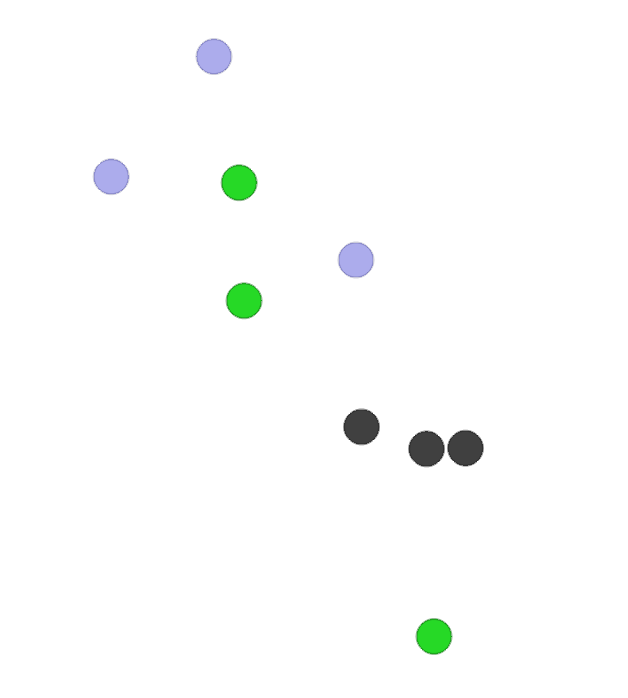}}}\quad
  \subfigure[Coverage]
  {\fbox{\includegraphics[height=0.22\linewidth]{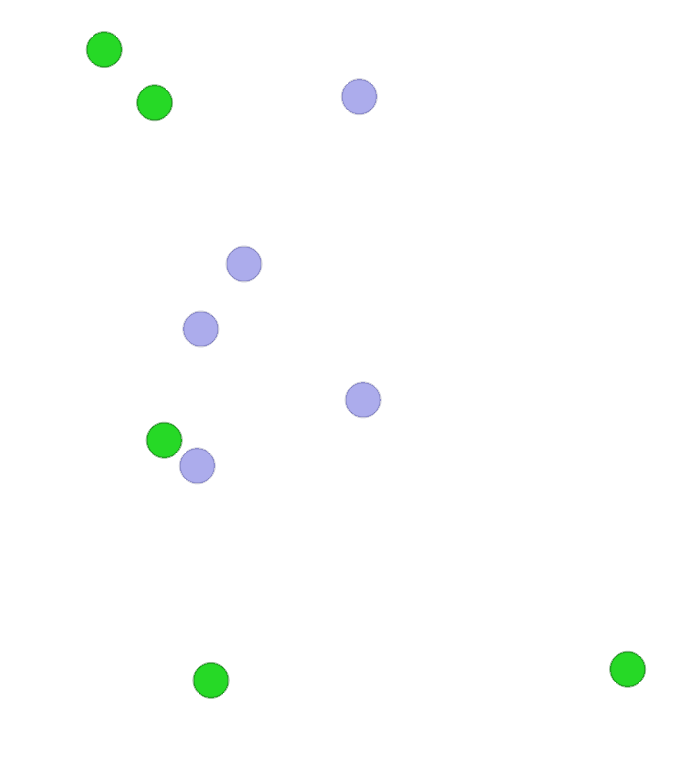}}}\quad
  \subfigure[Formation]
  {\fbox{\includegraphics[height=0.22\linewidth]{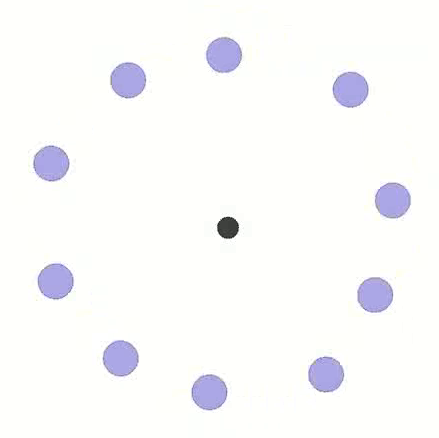}}}\quad
  \subfigure[Line]{\fbox{\includegraphics[height=0.22\linewidth]{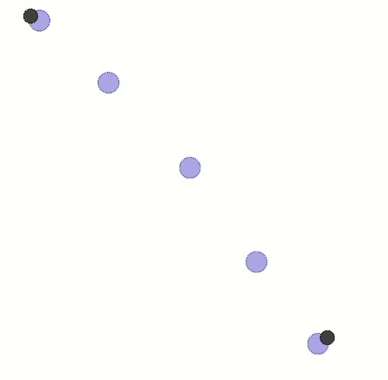}}}
  \label{figure:env}
  \caption{The agents are shown in blue circles, the goals are shown in green and obstacles are shown in black in the Target and Coverage environment. The landmarks are shown in black in the Formation and Line environments.}
\end{figure*}

\subsection{Target}
    There are $N$ agents, $N$ goals along with static obstacles in the environment. Each agent is supposed to go to its distinct goal while avoiding collisions with other entities in the environment. Agents start at random locations at the beginning of each episode; the corresponding goals are also randomly distributed. 
    Each agent $i$ gets a reward: $r_t^{(i)} = r_{\mathrm{dist},t}^{(i)} + r_{\mathrm{coll},t}^{(i)} + r_{\mathrm{goal},t}^{(i)}$, where $r_{\mathrm{dist}, t}^{(i)}$ is the negative of the Euclidean distance to the goal, $r_{\mathrm{coll},t}^{(i)}=-5$ if it collides with any other entity and zero otherwise, and $r_{\mathrm{goal},t}^{(i)}=+5$ if the agent has reached the goal and zero otherwise. The joint reward function is defined as $R(s_t,A_t)=\sum_{i=1}^{N}r_t^{(i)}$, which encourages cooperation among all agents.

\subsection{Coverage}
In the \emph{Coverage} environment \cite{targetCoverage1, targetCoverage2}, there are $N$ agents and $N$ goals in the environment. A major difference compared to the \emph{Target} environment is that each agent can go to any goal instead of going to a specific goal. The agents have to avoid collisions with other entities in the environment, and ensure that only one agent is present at each goal. 
To get the rewards for each agent, a linear sum assignment problem is solved at each time step where agents are assigned to goals depending on which one is the closest. The joint reward for all the agents is the negative of the mean of the minimum Euclidean distances to these assigned goals.

\subsection{Formation}
In the \emph{Formation} environment \cite{EMP}, there is a single landmark along with $N$ agents. The agents have to position themselves in an $N$-sided regular polygon with the landmark at its centre. The agents are rewarded at each time step according to how close they are to their expected positions. These expected positions are assigned by solving a linear sum assignment problem at each time step and depends on the number of agents in the environment and the desired radius of the polygon. We set the target radius to 0.5 for our experiments.

\subsection{Line} In the \emph{Line} environment, there are $N$ agents and two landmarks. The agents have to position themselves in an equally spread-out line between these two landmarks. Similar to the \emph{Formation} environment, the agents are rewarded according to how close they are to their expected positions. These expected positions are assigned by solving a linear sum assignment problem at each time step.

\section{Hyperparameters}
\label{appendix:hyperparams}
Tables \ref{table:informarl-hp}, \ref{table:mappo-hp}, \ref{table:offpolicy-hp}, \ref{table:all-hp} show the hyperparameters for InforMARL, MAPPO, MADDPG, MATD3, QMIX and VDN.

\begin{table*}[h]
  \centering
  \begin{tabular}{lc}
    \toprule
    hyperparameters     & Value \\
    \midrule
    entity embedding layer dim     &  3 \\
    entity hidden dim     &  16 \\
    num embedding layer     &  1 \\
    add self loop     &  False \\
    gnn layer hidden dim     &  16 \\
    num gnn heads     &  3 \\
    num gnn layers &  2\\
    gnn activation     &  ReLU \\
    \bottomrule
  \end{tabular}
  \caption{Hyperparameters used in InforMARL}
  \label{table:informarl-hp}
\end{table*}

\begin{table*}[h]
  \centering
  \begin{tabular}{lc}
    \toprule
    Common Hyperparameters     & Value \\
    \midrule
    recurrent data chunk length & 10 \\
    gradient clip norm     & 10.0 \\
    gae lambda     & 0.95 \\
    gamma     & 0.99 \\
    value loss     &  huber loss \\
    huber delta     &  10.0 \\
    batch size     &  num envs $\times$ buffer length $\times$ num agents\\
    mini batch size     &  batch size / mini-batch\\
    optimizer     &  Adam \\
    optimizer epsilon     &  1e-5 \\
    weight decay     &  0 \\
    network initialisation     &  Orthogonal \\
    use reward normalisation     &  True \\
    use feature normalisation     &  True \\
    \bottomrule
  \end{tabular}
  \caption{Common Hyperparameters used in MAPPO and InforMARL}
  \label{table:mappo-hp}
\end{table*}

\begin{table*}[h]
  \centering
  \begin{tabular}{lc}
    \toprule
    Common Hyperparameters     & Value \\
    \midrule
    gradient clip norm & 10.0 \\
    random episodes &  5 \\
    epsilon &  1.0 $\rightarrow$ 0.05\\
    epsilon anneal time &  50000 timesteps\\
    train interval & 1 episode \\
    gamma & 0.99 \\
    critic loss & mse loss \\
    buffer size & 5000 episodes \\
    batch size & 32 episodes \\
    optimizer & Adam \\
    optimizer eps & 1e-5 \\
    weight decay & 0 \\
    network initialisation & Orthogonal \\
    use reward normalisation & True \\
    use feature normalisation & True \\
    \bottomrule
  \end{tabular}
  \caption{Common Hyperparameters used in MADDPG, MATD3, QMIX, VDN}
  \label{table:offpolicy-hp}
\end{table*}

\begin{table*}[h]
  \centering
  \begin{tabular}{lc}
    \toprule
    Common Hyperparameters     & Value \\
    \midrule
    num envs     & 128 \\
    buffer length     &  25 \\
    num GRU layers     &  1 \\
    RNN hidden state dim     &  64 \\
    fc layer hidden dim     &  64 \\
    num fc     &  2 \\
    num fc after     &  1\\
    \bottomrule
  \end{tabular}
  \caption{Common Hyperparameters used in MAPPO, MADDPG, MATD3, QMIX, VDN and InforMARL}
  \label{table:all-hp}
\end{table*}

\section{Full Comparison \label{appendix:full_comparison}}
\begin{figure*}[h!]
\includegraphics[width=0.33\textwidth]{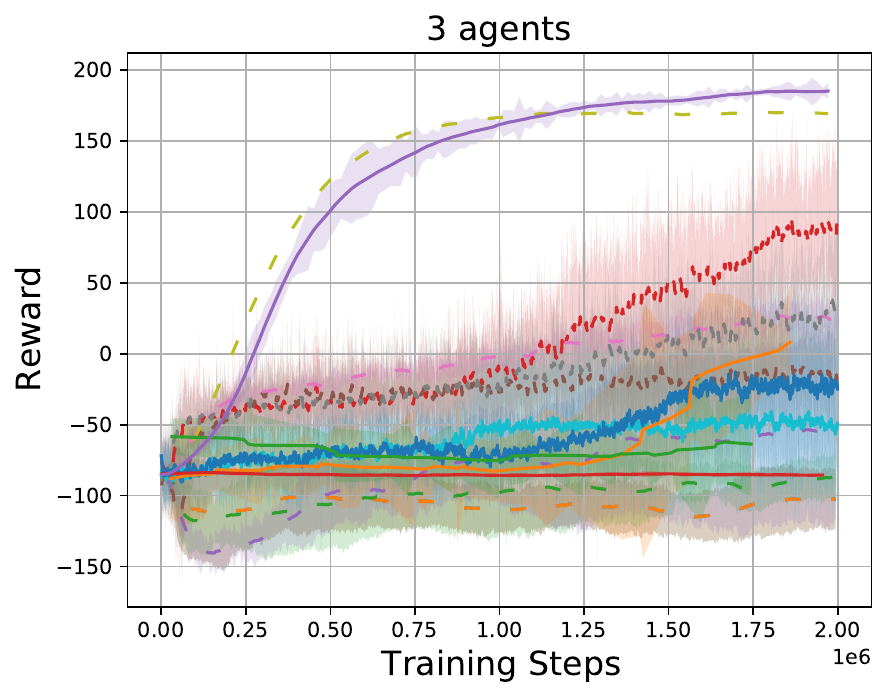}
  %
            \includegraphics[width=0.33\textwidth]{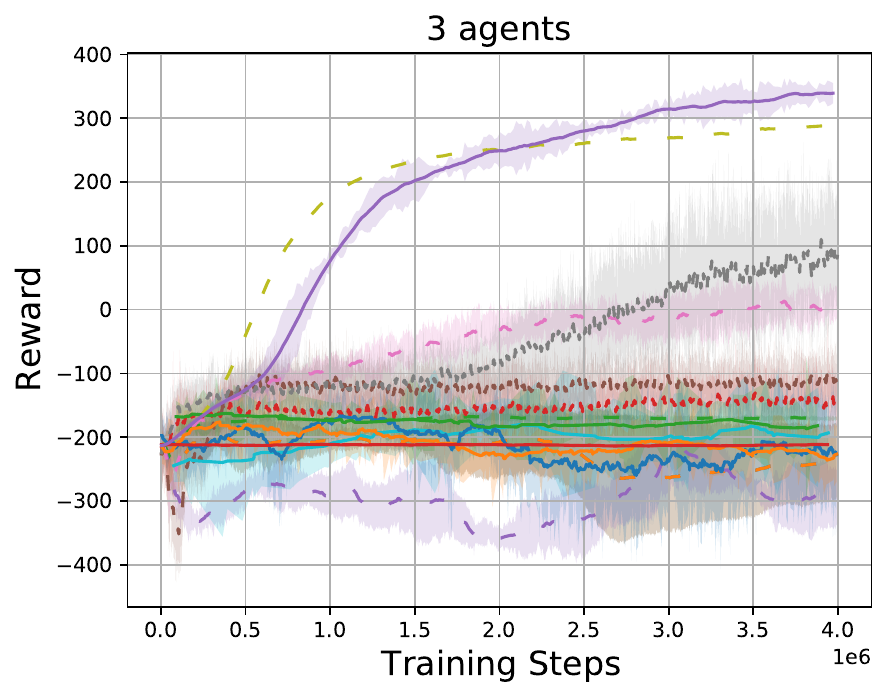}
 %
  \includegraphics[width=0.33\textwidth]{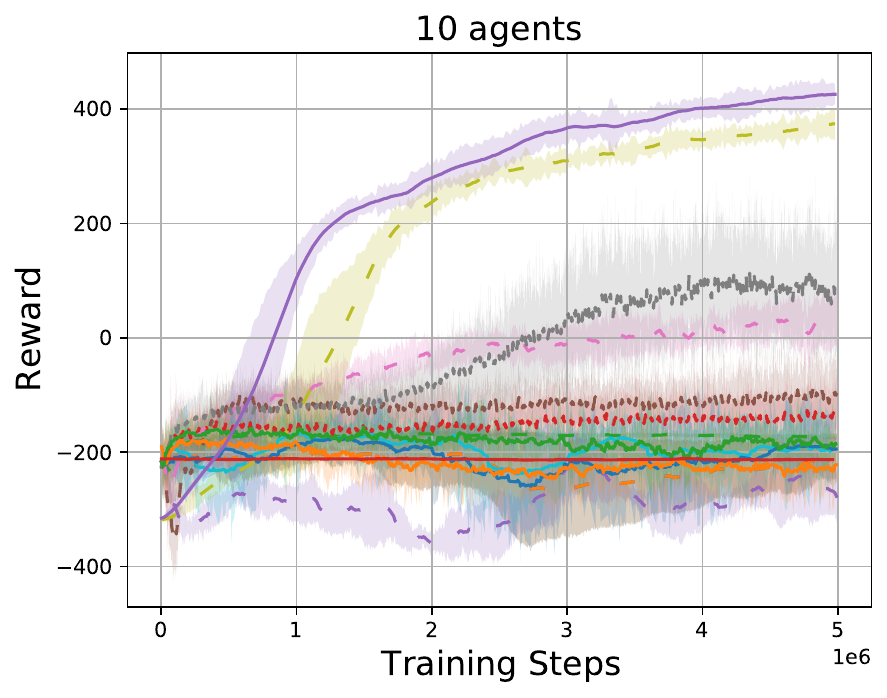}
 %
      \centering   \includegraphics[width=0.8\textwidth]{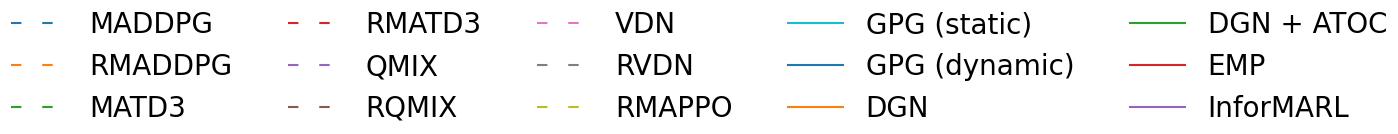}
        \caption{Comparison of InforMARL with baselines. The algorithms that use global information modes are represented with dashed lines and the algorithms that use local information modes are represented with solid lines.}
        \label{fig:compare_plots_long}
    \end{figure*}

We showcase the learning curves of all the baseline algorithms in this section in Figure \ref{fig:compare_plots_long}.

\begin{table*}
        \begin{tabular}{|p{1.8cm}||c||>{\centering}p{1.1cm}|>{\centering}p{0.5cm}|>{\centering}p{0.5cm}|>{\centering}p{0.35cm}||>{\centering}p{1.1cm}|>{\centering}p{0.5cm}|>{\centering}p{0.5cm}|>{\centering}p{0.35cm}||p{1.1cm}|>{\centering}p{0.5cm}|>{\centering}p{0.5cm}|c|}
        \hline
        \multirow{2}{*}{Algorithm} & Info & \multicolumn{4}{c||}{$N=3$} & \multicolumn{4}{c||}{$N=7$} & \multicolumn{4}{c|}{$N=10$}\\
        \cline{3-14}
         & mode & Reward & $T$ & \# col  & $S\%$ & Reward & $T$ & \# col & $S\%$ & Reward & $T$ & \# col & $S\%$\\
    \hline \hline
        MADDPG           & Global          &   $-100.73$     & $0.97$  &      $1.60$   &   5   &   $-206.07$     &  $0.98$ &    $6.01$      &   0  &  -210.43   &  1.00 &   9.26    &   0     \\ \hline
        RMADDPG          & Global          &   $10.73$     &  $0.75$ &         $1.72$   &  23 & $-122.07$  & $0.95$  & $4.61$ &  3 &     -127.98          &  1.00      &  7.39 &      0         \\ \hline
        MATD3             & Global          &   $-90.31$     &  $0.98$ &    $1.11$     & 5 &   $-169.08$     & $0.99$  &         $1.98$      & 1  &  -173.20   &  1.00 &   3.50 &    0       \\ \hline
        RMATD3           & Global          &   $105.49$     & $0.51$  &     $1.62$      &  67  &     $-128.93$   & $0.96$  &      $3.67$         &  6  &  -131.72  & 0.99   &  5.94  &    1       \\ \hline
        QMIX              & Global          &   $-54.24$     &  $0.84$ &         $0.71$      & 7  &  $-288.81$   & $0.97$  &     $3.92$     &  0    &     -273.46   & 1.00  & 5.98 &       0      \\ \hline
        RQMIX           & Global          &    $19.21$    &  $0.77$ &       $0.83$     & 28  &   $-83.41$     & $0.85$  &    $5.83$      & 12 &   -76.98     &  0.96 &  8.65  &     2      \\ \hline
        VDN              & Global          &   $18.86$     &  $0.67$ &        $1.56$       & 27 &  $39.87$    & $0.64$  &      $4.62$   &  23  &    43.23    &  0.73 &  5.79  &    19       \\ \hline
        RVDN             & Global          &    $64.04$    &  $0.62$ &          $0.57$     &  45  &  $140.94$  &  $0.62$ &       $3.42$   &   47  &    157.63    &  0.64 &  4.93  &  43           \\ 
        \specialrule{.15em}{.0em}{0em} 
            \textbf{RMAPPO}           & \textbf{Global}          &   \textbf{173.13}     & \textbf{0.41}  &        \textbf{0.67}   &  \textbf{96} &     \textbf{327.39}   & \textbf{0.44}  &      \textbf{4.29}    &  \textbf{88}    &  \textbf{366.81}      &  \textbf{0.44} &   \textbf{6.93}  &  \textbf{79}        \\ \specialrule{.15em}{.0em}{0em} 
        GPG (static)      & Local       &   $-67.03$     &  $0.96$ &      $0.72$     &  7  &  $-180.14$      & $0.99$  &      $3.27$     &   1 &    -182.57    & 1.00  &   4.28   &    0     \\ \hline
        GPG (dyn.)    & Local       &   $-46.27$     &  $0.87$ &     $0.21$     &   8  &  $-165.91$       & $1.00$  &        $1.57$       &  3  & -173.53  &  1.00 &   2.24   &    0     \\ \hline
        DGN               & Local       &    $32.94$    & $0.59$  &       $1.47$    &   32 &  $-232.32$      & $0.97$  &       $2.12$   &  0   &     -243.45   & 1.00  &   4.19  &     0     \\ \hline
        DGN+ATOC       & Local       &    $67.70$    &  $0.66$ &      $0.72$         &  35  &  $-189.61$  & $0.97$  &      $1.03$    &  0   &   -201.01     & 1.00  &   1.97   &  0       \\ \hline
        EMP              & Local       &  $-83.96$      &  $0.98$ &       $0.72$     & 6   &  $-211.90$      & $0.98$  &     $2.63$   &    0  & -209.90   & 1.00  &  94.12   &    0      \\ 
        \specialrule{.15em}{.0em}{0em} 
        \textbf{InforMARL} & \textbf{Local} &     \textbf{205.24}   & \textbf{0.17}  &       \textbf{1.45}    &  \textbf{100}  &     \textbf{399.01}   &  \textbf{0.37} &  \textbf{3.72}    &  \textbf{100}       &  \textbf{429.14}      &  \textbf{0.39} &  \textbf{4.73}   & \textbf{100}          \\ 
         \specialrule{.2em}{.0em}{0em} 
        \end{tabular}
        \caption{Comparison of InforMARL with other baseline methods, for scenarios with 3, 7, and 10 agents in the environment. The results presented represent the average of 100 test episodes. The following metrics are compared: (a) Total reward obtained in an episode by all the agents (higher is better). (b) Fraction of episode taken by the agents to reach the goal, \textbf{$T$} (lower is better). (c)   The total number of collisions the agents had in the episode, \# col (lower is better). (d) Percent of episodes in which all agents are able to get to their goals, $S$\% (higher is better). The best-performing methods that use global information (RMAPPO) and local information (InforMARL) are highlighted. As noted in Section \ref{sec:comparison_algos}, the metrics \# col and $S$ should be considered on balance.
        }
        \label{table:comparison_full}
    \end{table*}

\section{Ablation Studies}
\label{appendix:ablation}
    \subsection{Graph Information Aggregation Module}
\begin{figure}[h!]
        \centering
        \includegraphics[width=0.5\linewidth]{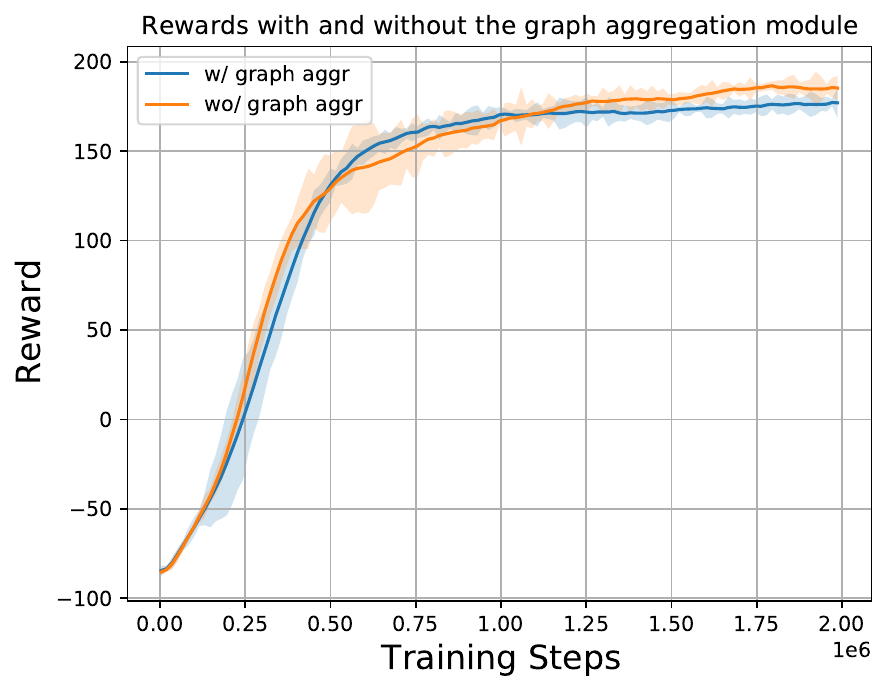}
        \caption{Training performance of InforMARL with, and without, the graph information aggregation module, for a 3-agent scenario. The two variants have similar sample complexities. However, the critic network with the graph information aggregation module has fewer parameters than the one without this module.}
        \label{fig:cent_obs}
    \end{figure}

    The graph information aggregation module allows our method to perform transfer learning to more complex environments. In this section, we compare models with and without the graph information aggregation module. In the absence of the graph information aggregation module, the states of individual agents are concatenated to be given as input to the centralized critic. Figure \ref{fig:cent_obs} presents the results of this ablation study. We find that both models have similar sample complexities and performance. However, the number of parameters for the critic network with the graph information aggregation module is much smaller and independent of the number of entities in the environment.
    
\subsection{Effect of Sensing Radius}
\begin{figure}[h!]
        \centering
        \includegraphics[width=0.5\linewidth]{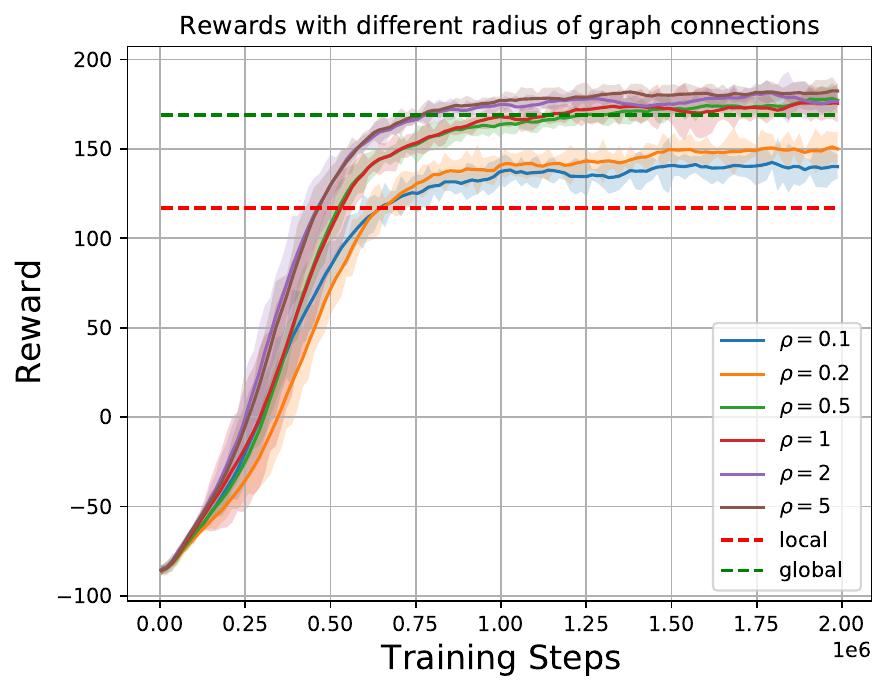}
        \caption{Diminishing returns in performance gains from increasing the sensing radius for InforMARL. The dashed lines are the reward values after saturation for RMAPPO in the global (in green) and local (in red) information modes, respectively. They are provided for reference.}
        \label{fig:compare_rad}
    \end{figure}

    We investigate how the performance of InforMARL depends on the sensing radius, namely, how much information (\ie over what neighborhood of the ego-agent) an agent has access to. 
    As the radius increases to a large value, the graph becomes fully connected; as the radius decreases to zero, the neighborhood information mode converges to the local information mode. As seen in Figure \ref{fig:compare_rad}, when learning with a small sensing radius (\eg $\rho=\{0.1, 0.2\}$), the agents are not able to achieve the same reward as can be achieved with a larger sensing radius ($\rho=\{0.5, 1, 2, 5\}$)\footnote{Measurements for $\rho$, a distance, are in meters}. We also note that there are diminishing returns when increasing the sensing radius from $\rho=0.5$ to $\rho=5$. Physically, a radius of $\rho=0.5$ is slightly more than double the distance an agent can traverse in the next two time steps, whereas $\rho=0.2$ is the distance it can travel in the current timestep. Since the agents far away from the ego-agent have little influence on their immediate decisions, the extra information obtained by increasing the sensing radius does not improve performance very much.

